\documentclass[aps,prl,twocolumn,
,floatfix,preprintnumbers,nofootinbib,superscriptaddress]{revtex4}

\usepackage{graphicx}
\usepackage{color}
\usepackage{natbib}
\usepackage{amsmath}
\usepackage{epstopdf}
\usepackage{hyperref}

\newcommand{\fref}[1]{Fig.~\ref{#1}}

\begin{document}

\title{A note of clarification: BICEP2 and Planck are not in tension}

\author{Benjamin Audren}
\affiliation{Institut de
Th\'eorie des Ph\'enom\`enes Physiques, \'Ecole Polytechnique
F\'ed\'erale de Lausanne, CH-1015, Lausanne,
Switzerland}

\author{Daniel G. Figueroa}
\affiliation{D\'epartement de Physique Th\'eorique and Center for Astroparticle Physics,
Universit\'e de Gen\`eve, 24 quai Ernest Ansermet, CH1211 Gen\`eve 4, Switzerland}

\author{Thomas Tram}
\affiliation{Institut de
Th\'eorie des Ph\'enom\`enes Physiques, \'Ecole Polytechnique
F\'ed\'erale de Lausanne, CH-1015, Lausanne,
Switzerland}

\date{\today}


\begin{abstract}
The apparent discrepancy between the value of the tensor-to-scalar ratio
reported by the BICEP2 collaboration, $r = 0.20^{+0.07}_{-0.05}$ at 68\% CL, and the
Planck upper limit, $r < 0.11$ at 95\% CL, has attracted a great deal of
attention. In this short note, we show that this discrepancy is mainly due to
an `apples to oranges' comparison. The result reported by BICEP2 was measured
at a pivot scale $k_* = 0.05$ Mpc$^{-1}$, assuming $n_t = 0$, whereas the
Planck limit was provided at $k_* = 0.002$ Mpc$^{-1}$, assuming the slow-roll
consistency relation $n_t = -r/8$. One should obviously compare the BICEP2 and
Planck results under the same circumstances.  By imposing $n_t = 0$, the Planck
constraint at $k_* = 0.05$ Mpc$^{-1}$ becomes $r < 0.135$ at $95\%$ CL, which
can be compared directly with the BICEP2 result.
Once a plausible dust contribution to the BICEP2 signal is taken into account (DDM2 model),
$r$ is reduced to $r = 0.16^{+0.06}_{-0.05}$ and the discrepancy becomes of
order $1.3\sigma$ only.
\end{abstract}

\maketitle

\section{The pivot scale confusion}\label{sec:pivots}

\begin{figure}[t]%
\includegraphics[width=\columnwidth]{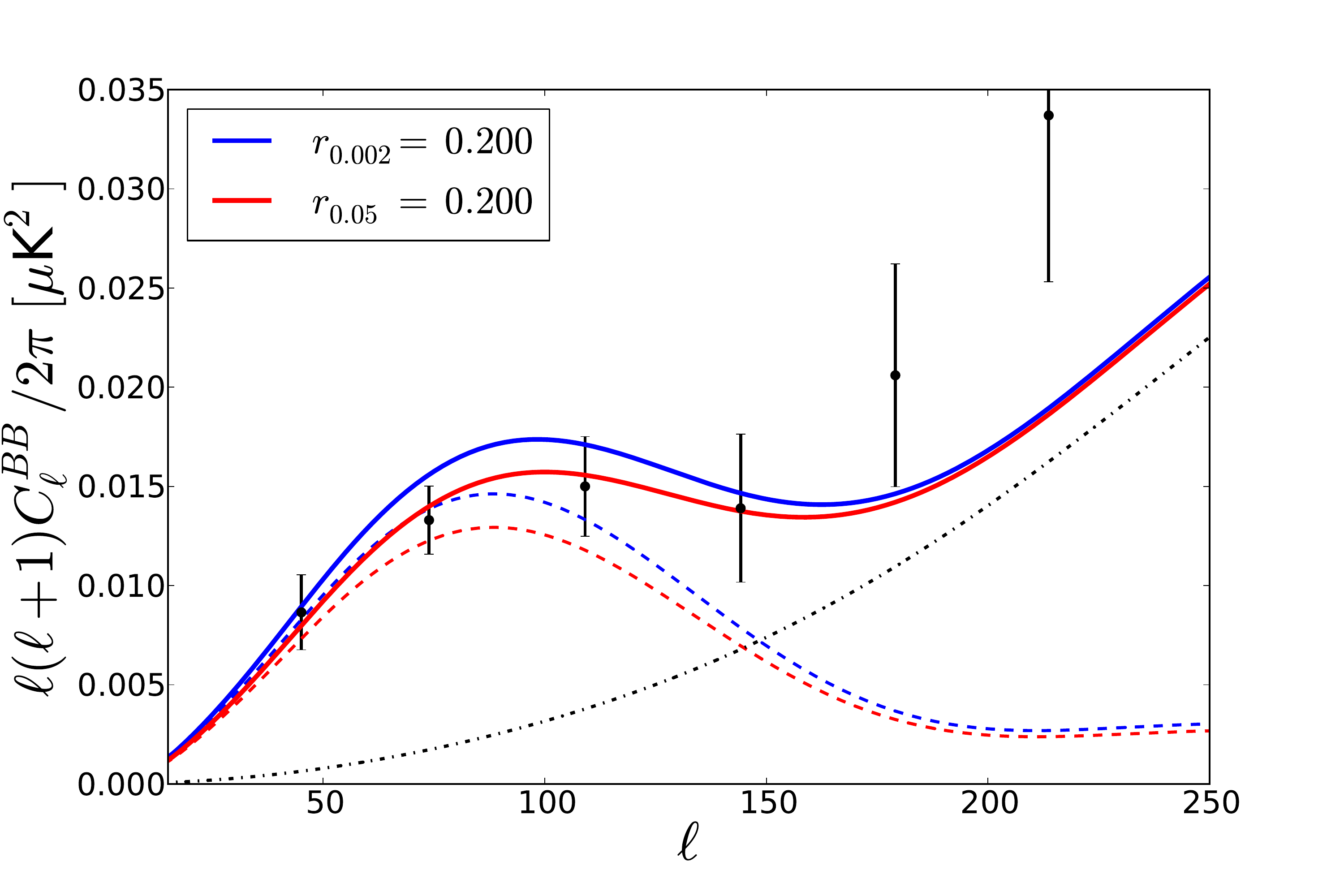}%
\caption{B-mode angular power spectra for $r_{0.05} = 0.2$ (red curves) and
$r_{0.002} = 0.2$ (blue curves), both for $n_t = \alpha_t = \alpha_s = 0$. The
data points and error bars are from the table available at~\url{bicepkeck.org}, and
we have assumed Planck best-fit values for other cosmological parameters, in
particular $A_s=2.215\times 10^{-9}$ and $n_s=0.9619$. The dashed lines
correspond to the B-modes from inflationary tensors, the dotted-dashed line
show the contribution from the lensing of the (predominantly scalar) E-modes,
and the continuous lines are the sum of the two.}
\label{fig:BicepFit}%
\end{figure}

A generic prediction of the inflationary paradigm is that the primordial scalar
and tensor power spectra from inflation are nearly scale invariant. The
deviation from scale invariance is then quantified by specifying a spectral
index $n$ and possibly a running $\alpha$ at a pivot scale $k_*$. The
primordial power spectra are then given by
\begin{align}
\mathcal{P}_s(k) &= A_s \left( \frac{k}{k_*} \right)^{(n_s-1) + \frac{1}{2} \alpha_s  \log \left(\frac{k}{k_*} \right) } \label{eq:scalar}\\ 
\mathcal{P}_t(k) &= A_t \left( \frac{k}{k_*} \right)^{n_t+\frac{1}{2} \alpha_t \log \left(\frac{k}{k_*} \right)  } \label{eq:tensor}
\end{align}
where $\alpha_s \equiv dn_s/d\log k$ and $\alpha_t \equiv dn_t/d\log k$. The
primordial tensor amplitude $A_t$ is, by convention, always substituted for the
ratio
\begin{equation}
r \equiv \frac{\mathcal{P}_t(k_r)}{\mathcal{P}_s(k_r)}\,,
\end{equation}
evaluated at a given scale $k_r$, which is often, but not always, taken to be
$k_*$. The analysis of the temperature anisotropies by the Planck
collaboration~\cite{Ade:2013zuv} was done at a pivot scale $k_* =
0.05~\text{Mpc}^{-1}$, but their reported constraint $r < 0.11$ at $95\%$ Confidence Level (CL)
was given at $k_r = 0.002~\text{Mpc}^{-1}$. The BICEP2
collaboration~\cite{Ade:2014xna}, on the other hand, reported the amplitude $r
= 0.20^{+0.07}_{-0.05}$ at 68\% CL, evaluated at the Planck pivot scale\footnote{Note that
this is not explicitly stated in~\cite{Ade:2014xna}, at least not in the current
arXiv version at the time of writing this note, but it has been confirmed to us
by the BICEP2 collaboration.} $k_r = k_* = 0.05\, \text{Mpc}^{-1}$. To avoid
any confusion, it is then convenient to denote the tensor-to-scalar ratios
evaluated at $k_r = 0.05\,\text{Mpc}^{-1}$ and $k_r = 0.002\,\text{Mpc}^{-1}$
by $r_{0.05}$ and $r_{0.002}$ respectively. That BICEP2 indeed has
$r_{0.05}=0.2$ as best fit becomes evident in~\fref{fig:BicepFit}, where the
low-$\ell$ B-mode angular power spectrum are plotted for $r_{0.05} = 0.2$ (red
lines) and $r_{0.002} = 0.2$ (blue lines), considering for both cases $n_t =
0$, and assuming $\alpha_s = \alpha_t = 0$. The $r_{0.002} = 0.2$ curve is
clearly not a good fit to the data points given by BICEP2. 

\section{The likelihood confusion}

The confusion related to the scale $k_r$ is enhanced by the following
circumstances: {\it i)} the BICEP2 collaboration used a different likelihood in their own
analyses than the publicly released Python likelihood code, and {\it ii)} the
best-fit of the public code is actually $r_{0.002} = 0.2$, and \emph{not}
$r_{0.05} = 0.2$ as found in the BICEP2 analysis. This difference in best-fit
is due to two separate facts: {\it a)} different methods are being used for
computing the likelihood, the public one using the Hamimeche \& Lewis code,
whereas the private one uses the formula introduced in~\cite{Barkats:2013jfa},
paragraph 9.3.1, and {\it b)} the public code uses information from all nine
bandpower bins, whereas the internal one makes use of only the five first ones.

\begin{figure}[t]%
\includegraphics[width=\columnwidth]{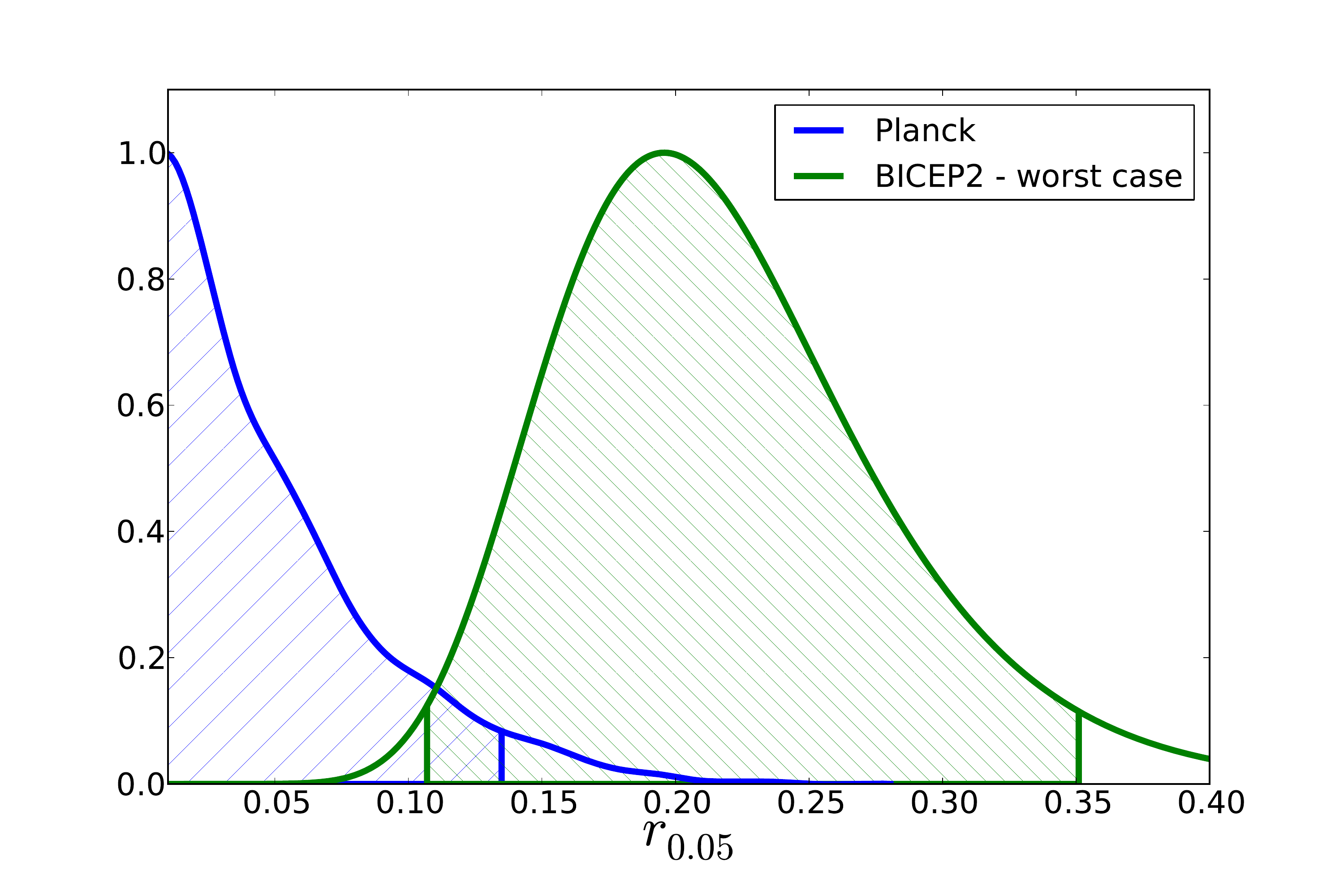}\\
\includegraphics[width=\columnwidth]{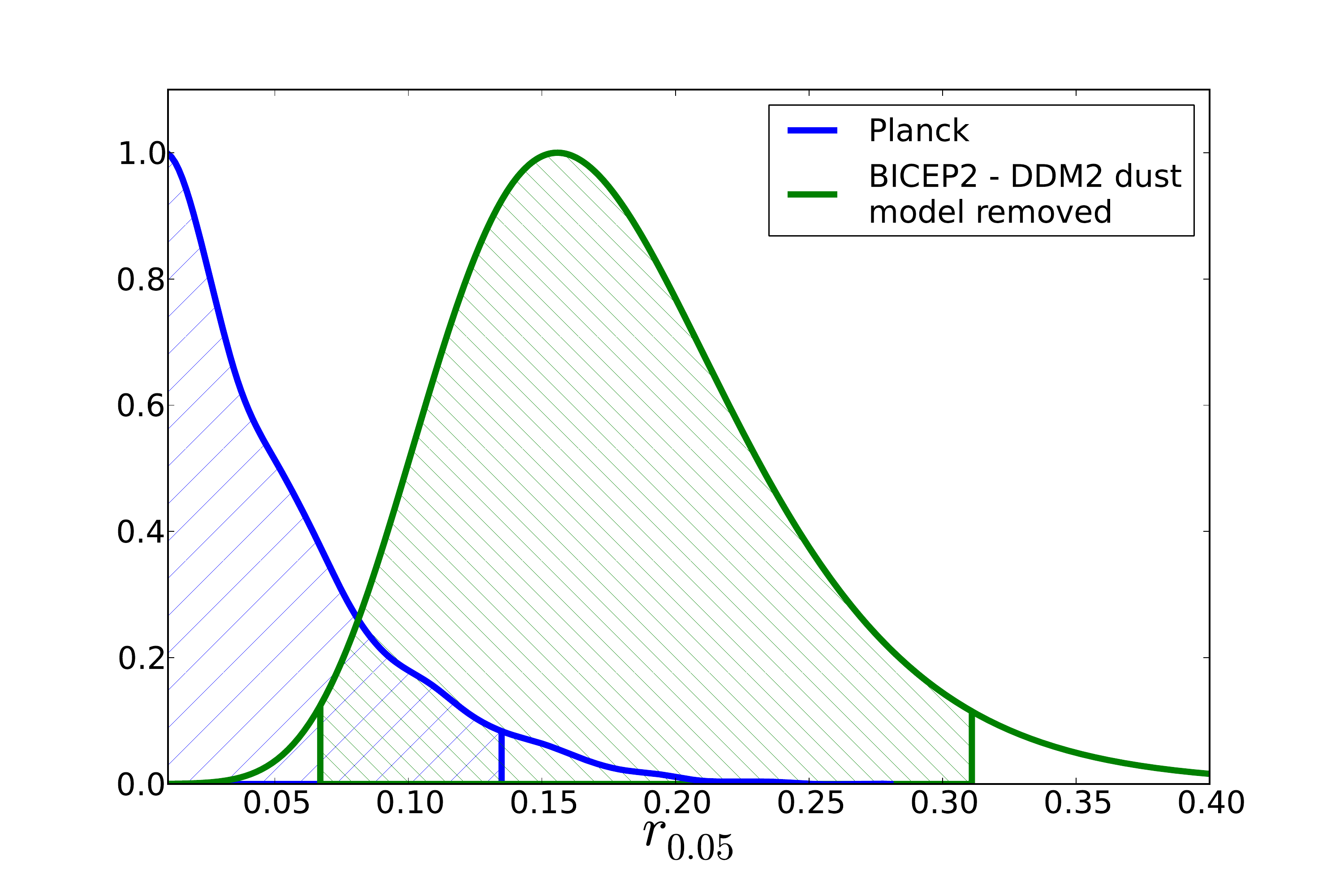}
\caption{Posteriors for $r_{0.05}$ from both Planck (blue) and BICEP2 (green). For BICEP2,
in the top panel, we plot the tabulated likelihood available from \url{bicepkeck.org}, obtained assuming Planck
best-fit values for the remaining cosmological parameters; in the bottom panel, we shift the same distribution by $-0.04$,
to account approximatively for dust removal assuming the plausible DDM2 model discussed by the collaboration.
For Planck, we show the
full posterior distribution obtained with a Markov Chain Monte Carlo sampling of
the 6 Standard Model parameters and $r_{0.05}$, as well as the 14 nuisance
parameters of Planck, assuming $n_t = \alpha_t =\alpha_s = 0$. The vertical lines encapsulate
the $2\sigma$ allowed regions for each distribution.}\label{fig:posteriors}
\end{figure}

It should be noted that the difference between the best-fit values of the two
likelihoods is well below $1 \sigma$. So this is not alarming in any way, but it
leads nonetheless to an overestimation of the tension with Planck when using
the public code.

In conclusion, the only data product matching exactly the BICEP2
internal analysis is the tabulated likelihood, obtained for a fixed cosmology with different
values of $r_{0.05}$, represented in green in the top panel of
Fig.~\ref{fig:posteriors}. This corresponds to the advertised value
of $r_{0.05} = 0.20^{+0.07}_{-0.05}$. Reference~\cite{Ade:2014xna} also discusses several dust models,
retaining DDM2 (Data Driven Model 2) as the most plausible one. After removing the DDM2 contamination,
the BICEP2 collaboration obtains $r_{0.05} = 0.16^{+0.06}_{-0.05}$ (68\% CL). 
In the lower panel of Fig.~\ref{fig:posteriors} we present an approximative $r$ posterior after dust removal.


\section{Comparison with Planck}

Using Eqs.~\eqref{eq:scalar}~and~\eqref{eq:tensor}, one can convert $r_{0.05}$
from BICEP2 to $r_{0.002}$ for comparison with Planck, or vice versa. For
instance,  the $r_{0.05} = 0.2$ curve in~\fref{fig:BicepFit}, which is the best
fit to the data (for $n_t = \alpha_t = \alpha_s = 0$), is equivalent to
$r_{0.002} \simeq 0.177$. However, this would still be an `apples to oranges'
comparison, since the Planck analysis used a tensor spectral index inferred
from the single-field slow-roll consistency condition $n_t = -r/8$, while
BICEP2 used $n_t = 0$. This means that the underlying tensor primordial spectra
was not of the same form, so it is in principle meaningless to compare the two
parameters: If one experiment fits $y = a_0 + a_1 x$ to the data while the
other fits $y = b_0$, we certainly should not compare $a_0$ and $b_0$. 

We derived the posterior probability for $r_{0.05}$ assuming a flat
$\Lambda\text{CDM}+r$ model and the Planck+WP dataset. In any Bayesian
parameter extraction, the posterior depends on the choice of prior. Here, we
choose to restrict ourselves to physical models by imposing a prior $r_{0.05}
\geq 0$ (a different choice is advocated in  the recent analysis of
\cite{Smith:2014kka}). After running the {\sc class} and {\sc Monte Python}
codes, we obtained $r_{0.05} < 0.135$ at 95\% CL, which is not in significant
tension with the BICEP2 result, as shown in~\fref{fig:posteriors}.  Even before
subtracting the dust model, the two posteriors overlap at the 9\% CL
(corresponding to 1.7$\sigma$). After removing dust contamination (under the
DDM2 assumption), the compatibility increases\footnote{Here the compatibility
is quantified by searching for the confidence level of each likelihood above
which there is an overlap. Another statistical test of the compatibility
between two such likelihoods is presented in \cite{Smith:2014kka}.} to the
level of 17\%, corresponding to a 1.3$\sigma$ overlap.

With such an overlap between the two likelihoods, we can conclude (even without
calculating Bayesian evidence ratios) that there is no compelling reason at the
moment to invoke extra ingredients in the cosmological model, in order to
alleviate a would-be tension between the Planck 2013 and BICEP2 measurements.
In particular, there is no convincing case for introducing a non-zero scalar
running $\alpha_s$ of the order of $-0.02$, which would be incompatible with
the simplest and most elegant slow-roll inflationary paradigm.



\section*{Acknowledgements}
We would like to thank Clem Pryke and Stefan Fliescher from the BICEP2 collaboration for their prompt and very helpful answers to all our questions.

\bibliography{BICEP2vsPlanck_Arxiv1}

\begin{thebibliography}{4}
\expandafter\ifx\csname natexlab\endcsname\relax\def\natexlab#1{#1}\fi
\expandafter\ifx\csname bibnamefont\endcsname\relax
  \def\bibnamefont#1{#1}\fi
\expandafter\ifx\csname bibfnamefont\endcsname\relax
  \def\bibfnamefont#1{#1}\fi
\expandafter\ifx\csname citenamefont\endcsname\relax
  \def\citenamefont#1{#1}\fi
\expandafter\ifx\csname url\endcsname\relax
  \def\url#1{\texttt{#1}}\fi
\expandafter\ifx\csname urlprefix\endcsname\relax\def\urlprefix{URL }\fi
\providecommand{\bibinfo}[2]{#2}
\providecommand{\eprint}[2][]{\url{#2}}

\bibitem[{\citenamefont{Ade et~al.}(2013)}]{Ade:2013zuv}
\bibinfo{author}{\bibfnamefont{P.}~\bibnamefont{Ade}} \bibnamefont{et~al.}
  (\bibinfo{collaboration}{Planck Collaboration}) (\bibinfo{year}{2013}),
  \eprint{1303.5076}.

\bibitem[{\citenamefont{Ade et~al.}(2014)}]{Ade:2014xna}
\bibinfo{author}{\bibfnamefont{P.}~\bibnamefont{Ade}} \bibnamefont{et~al.}
  (\bibinfo{collaboration}{BICEP2 Collaboration}) (\bibinfo{year}{2014}),
  \eprint{1403.3985}.

\bibitem[{\citenamefont{Barkats et~al.}(2013)}]{Barkats:2013jfa}
\bibinfo{author}{\bibfnamefont{D.}~\bibnamefont{Barkats}} \bibnamefont{et~al.}
  (\bibinfo{collaboration}{BICEP1 Collaboration}) (\bibinfo{year}{2013}),
  \eprint{1310.1422}.

\bibitem[{\citenamefont{Smith et~al.}(2014)\citenamefont{Smith, Dvorkin, Boyle,
  Turok, Halpern et~al.}}]{Smith:2014kka}
\bibinfo{author}{\bibfnamefont{K.~M.} \bibnamefont{Smith}},
  \bibinfo{author}{\bibfnamefont{C.}~\bibnamefont{Dvorkin}},
  \bibinfo{author}{\bibfnamefont{L.}~\bibnamefont{Boyle}},
  \bibinfo{author}{\bibfnamefont{N.}~\bibnamefont{Turok}},
  \bibinfo{author}{\bibfnamefont{M.}~\bibnamefont{Halpern}},
  \bibnamefont{et~al.} (\bibinfo{year}{2014}), \eprint{1404.0373}.

\end{thebibliography}


\end{document}